\documentclass[11pt]{article}
\usepackage{CJKnumb}
\usepackage{amsmath, amsfonts, bm,cite}
\usepackage{newtxtext} % Times New Roman text
\usepackage{newtxmath} % Times New Roman math
\usepackage[pdftex]{graphicx}
\usepackage{color}
\usepackage{multirow}
\usepackage{fancyhdr}
\usepackage{array}
\usepackage{makeidx, chngpage}
\usepackage{enumerate}
\usepackage{titlesec}
\usepackage[T1]{fontenc}
\usepackage{float}
\usepackage{url}
\usepackage{caption}
\usepackage{booktabs}
\usepackage{ascii}
\usepackage[margin=1.5cm]{geometry}
\usepackage{bm}
\usepackage{indentfirst}
\usepackage{textcomp}
\usepackage{graphicx}
\usepackage{hyperref} % Enable hyperlinks
\hypersetup{
    colorlinks=true,        % Enable colored links
    linkcolor=blue,         % Color for internal links (e.g., equations, sections)
    citecolor=blue,          % Color for citations
    urlcolor=purple         % Color for URLs
}
\usepackage{mwe}
\usepackage{subfigure}
\usepackage{appendix}
\usepackage{mathtools}

\usepackage{amstext}
\usepackage{fancyhdr}
\usepackage{lineno}
\nolinenumbers

\RequirePackage[table]{xcolor} % To provide color for soul (for english editing) and provide coloring for tables (author request)
\RequirePackage{soul} % To highlight text
\RequirePackage{multirow}

\newlength{\defbaselineskip}
\newcommand{\bitem}{\begin{itemize}} 
\newcommand{\eitem}{\end{itemize}}
\newcommand{\be}{\begin{equation}}
\newcommand{\ee}{\end{equation}}
\newcommand{\benu} {\begin{enumerate}}
\newcommand{\eenu} {\end{enumerate}}
\newcommand{\ba}{\begin{eqnarray}}
\newcommand{\ea}{\end{eqnarray}}
\newcommand{\bdes}{\begin{description}}
\newcommand{\edes}{\end{description}}

\title{Adaptive Digital Twin of Sheet Metal Forming via Proper Orthogonal Decomposition-Based Koopman Operator with Model Predictive Control}

\author{Yi-Ping Chen$^{1,a}$, Derick Suraez$^{1,a}$, Ying-Kuan Tsai$^{1}$, Vispi Karkaria$^{1}$, Guanzhong Hu$^{1}$ , Zihan Chen$^{1}$, \\ Ping Guo$^{1}$ , Jian Cao$^{1}$ , and Wei Chen$^{1}$\thanks{Corresponding author, weichen@northwestern.edu}\\ \small{$^{1}$ Department of Mechanical Engineering, Northwestern University} \\
\small{$^{a}$ Equally contributed}}
\date{November 13, 2025}

\begin{document}

% \nomenclature{AM}{Additive Manufacturing}
% \nomenclature{$d_x, d_y$}{Distances from the current laser position to the nearest geometry boundary along the $x-$ and $y-$ axis}
% \nomenclature{DED}{Directed Energy Deposition}
% \nomenclature{DNN}{Deep Neural Network}
% \nomenclature{NN}{Neural Network}
% \nomenclature{$\hat{f}(\cdot)$}{Surrogate model of the system}
% \nomenclature{$J(\cdot)$}{Objective function for MPC}
% \nomenclature{MPC}{Model Predictive Control}
% \nomenclature{$p$}{Horizon length}
% \nomenclature{TiDE}{Time Series Dense Encoder}
% \nomenclature{$u_{k}$}{Control input at timestep $k$}
% \nomenclature{$w$}{Window size}
% \nomenclature{$x_k$}{Actual state at timestep $k$}
% \nomenclature{$\hat{x}_k$}{Predicted state at timestep $k$}
% \nomenclature{$z$}{$z-$coordinate of the laser nozzle position}

\maketitle

\begin{abstract}
Digital Twin (DT) technologies are transforming manufacturing by enabling real-time prediction, monitoring, and control of complex processes. Yet, applying DT to deformation-based metal forming remains challenging because of the strongly coupled spatial-temporal behavior and the nonlinear relationship between toolpath and material response. For instance, sheet-metal forming by the English wheel, a highly flexible but artisan-dependent process, still lacks digital counterparts that can autonomously plan and adapt forming strategies. This study presents an adaptive DT framework that integrates Proper Orthogonal Decomposition (POD) for physics-aware dimensionality reduction with a Koopman operator for representing  nonlinear system in a linear lifted space for the real-time decision-making via model predictive control (MPC). To accommodate evolving process conditions or material states, an online Recursive Least Squares (RLS) algorithm is introduced to update the operator coefficients in real time, enabling continuous adaptation of the DT model as new deformation data become available. The proposed framework is experimentally demonstrated on a robotic English Wheel sheet metal forming system, where deformation fields are measured and modeled under varying toolpaths. Results show that the adaptive DT is capable of controlling the forming process to achieve the given target shape by effectively capturing non-stationary process behaviors. Beyond this case study, the proposed framework establishes a generalizable approach for interpretable, adaptive, and computationally-efficient DT of nonlinear manufacturing systems, bridging reduced-order physics representations with data-driven adaptability to support autonomous process control and optimization.
\end{abstract}

\tableofcontents

\section{Introduction}

The concept of the Digital Twin (DT) has opened new opportunities for real-time monitoring, prediction, and decision-making \cite{karkaria2024optimization} across industries such as aerospace \cite{kapteyn2021probabilistic}, energy \cite{yu2022energy}, and manufacturing \cite{abadi2025leveraging, karkaria2024towards}. By providing a virtual replica of a physical system, a DT enables in silico simulation of “what-if” scenarios and the optimization of control strategies through techniques such as model predictive control (MPC) \cite{chen2025real, chen2026uncertainty} or reinforcement learning (RL) \cite{tsai2025digital}. In manufacturing, particularly in metal forming processes characterized by complex, spatio-temporal, and multi-physics behaviors, machine learning (ML) approaches has been widely adopted to surrogate the high-fidelity simulations as it is nearly impossible to run in real-time \cite{gunasegaram2024machine, karkaria2025attention}. However, directly applying ML to these systems remains challenging: data are expensive to obtain, system responses are spatially and temporally coupled, and purely data-driven models often lack physical interpretability and generalizability beyond the training domain. These challenges highlight the need for model representations that embed physical principles while enabling data-driven learning and control.

Sheet metal forming remains ubiquitous worldwide, with an estimated market value exceeding \$480 billion USD and expected robust growth in the coming decade \cite{MarketResearchFuture2025SheetMetal}. Among forming techniques, traditional forming methods such hammering, shrinking/stretching and spinning (among others) have gained recent attention \cite{bowen2022} due to their high flexibility \cite{yang2018}, with recent advances in automated spinning and forging \cite{bandar2025}. Of the traditional forming tools, the English wheel, an asymmetric rolling process that locally stretches and deforms sheet metal, plays a unique role in producing smooth, multi-curvature panels lucrative to the aerospace, automotive, architecture and restoration industries. Traditionally, English wheel forming relies on skilled artisans to manually and iteratively adjust the applied compressive force and driving path/toolpath based on visual and tactile feedback. As the need for flexible manufacturing for custom high-quality parts increases and skilled manual manufacturing labor declines globally, it becomes increasingly urgent to digitalize the process.

\begin{figure*}[b]
    \centering
    \includegraphics[width=1\linewidth]{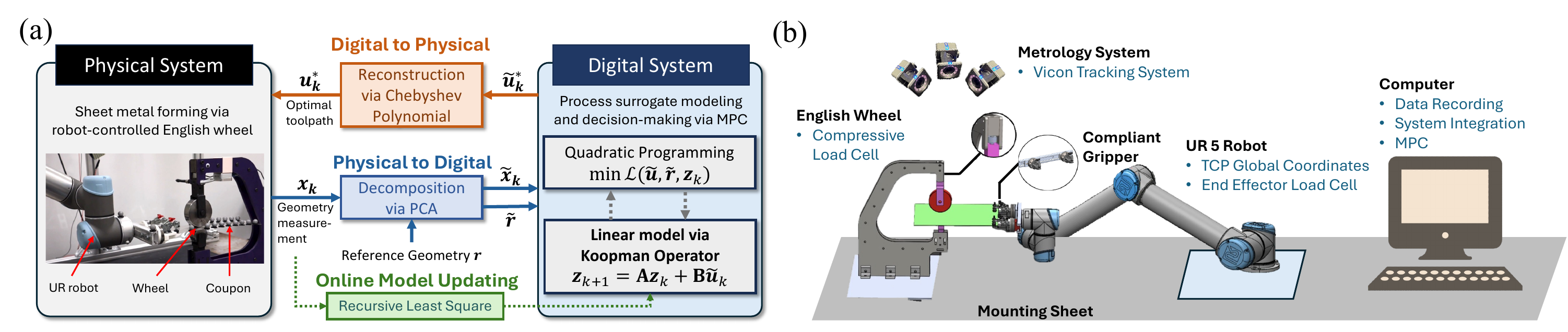}
    \caption{(a) The proposed framework for the dynamic process DT. (b) System setup for the robotic English wheel.}
    \label{fig:overall_flowchart}
\end{figure*}

In our previous work \cite{suarez2024}, we developed a novel robotic English wheel system for sheet metal forming that integrates load sensing, robotic actuation using a Universal Robot (UR), and real-time geometry tracking via a Vicon optical metrology system \cite{zhang2024process}, as illustrated in Figure \ref{fig:overall_flowchart}(b). While this system enables real-time feedback of the evolving sheet deformation, a critical challenge on real-time toolpath design remains unresolved. Specifically, it is still unclear \emph{how to systematically generate toolpaths that achieve a desired target geometry when both the control input (toolpath) and the system state (sheet deformation) are high-dimensional, strongly spatially coupled fields with nonlinear dynamics}. This challenge lies at the core of enabling autonomous, feedback-driven forming, as it requires a model that can accurately capture the spatio-temporal deformation behavior while remaining computationally tractable for real-time decision-making.

Prior studies have explored various toolpath design and control strategies for sheet metal forming to achieve desired product geometries under different target shapes, operational conditions, and variable material properties. Feedback control methods, such as PID controllers, have demonstrated significant success but are generally limited to single-input single-output systems \cite{allwood2016}, making them less suitable for spatially correlated input/output fields. RL has also gained attention for handling high-dimensional states \cite{liu2020reinforcement}, while its data-intensive nature makes it impractical for most forming scenarios where data acquisition is costly. Model-based approaches, such as finite element simulations, while accurate, are computationally prohibitive for online optimization and iterative design evaluation \cite{hou2023review}. Data-driven methods, mostly represent the incremental model \cite{lu2016model}, offer real-time feasibility for optimization-based control but often fail to capture the nonlinear deformation dynamics for global modeling \cite{wang2023learning, he2020switched}. These limitations motivate the development of a nonlinear surrogate model that can be seamlessly integrated with efficient control solvers to realize the DT of the process.

% \begin{figure}[b]
%     \centering
%     \vspace{5pt}
%     \includegraphics[width=\linewidth]{Figure/feedback_loop.pdf}
%     \caption{Closed-loop control methodology applied to the English wheel}
%     \label{fig:closed loop}
% \end{figure}

In the past decades, the Koopman operator emerges as a powerful framework for representing nonlinear dynamical systems through an infinite-dimensional but linear operator acting on observable functions \cite{brunton2016koopman, brunton2019notes}. When approximated in finite dimensions using data-driven approaches, such as extended dynamic mode decomposition (eDMD) , or Deep Neural Networks (DNN), the Koopman operator enables the application of well-established linear control techniques to nonlinear systems. This concept has demonstrated remarkable success in robotics and autonomous systems, including drone trajectory control \cite{abraham2019active} and soft robot manipulation \cite{bruder2020data}. Recent advances have extended Koopman-based modeling to manufacturing, for example in wire-arc additive manufacturing (WAAM), where PointNet-based observables are used to represent the evolution of deposited geometry \cite{biehler2024retrofit}. However, applying Koopman operators to sheet metal forming remains unexplored, particularly where both system states and control inputs are spatially correlated vector fields. Inspired by studies that combine Proper Orthogonal Decomposition (POD) for spatial feature extraction with Koopman representations for temporal evolution \cite{verma2025efficient, semeraro2012analysis}, we explore a spatial-temporal learning approach to enable a physics-aware DT for forming processes.

In this work, we propose a DT framework using adaptive POD-based deep Koopman operator with control for the DT of robotic English wheel sheet forming, as illustrated in Figure~\ref{fig:overall_flowchart}. First, POD and Chebyshev polynomial decomposition are applied to reduce the dimensionality of deformation and toolpath fields, respectively, while preserving their dominant spatial structures. A deep Koopman operator with residual-connected encoder-decoder architecture is then trained to learn the evolution of the reduced-order system. The learned linearized model is subsequently integrated with linear constrained MPC to generate optimal toolpaths that achieve a given target geometry in real time. To further mitigate the effects of data sparsity and evolving material behavior, an online model updating mechanism based on Recursive Least Squares (RLS) is developed to adaptively refine the control matrix during operation. The complete framework is validated experimentally using a fully automated robotic English wheel system, demonstrating the effectiveness and feasibility of the proposed DT-based approach. The key contributions of this work are summarized as follows:
\begin{itemize}
\item A DT integrating POD and deep Koopman operator for forming processes, effectively capturing spatial–temporal dynamics in a reduced-order representation.
\item A real-time decision-making framework combining Koopman-based linear modeling with constrained MPC for optimal toolpath design.
\item An online model adaptation mechanism using RLS to maintain model accuracy against material variability and model uncertainty.
\item Experimental validation of the proposed framework in a fully autonomous, closed-loop robotic English wheel system, demonstrating the feasibility of DT-enabled forming.
\end{itemize}

\section{System Overview and Experimental Procedure}

\begin{figure}[t]
    \centering
    \includegraphics[width=1\linewidth]{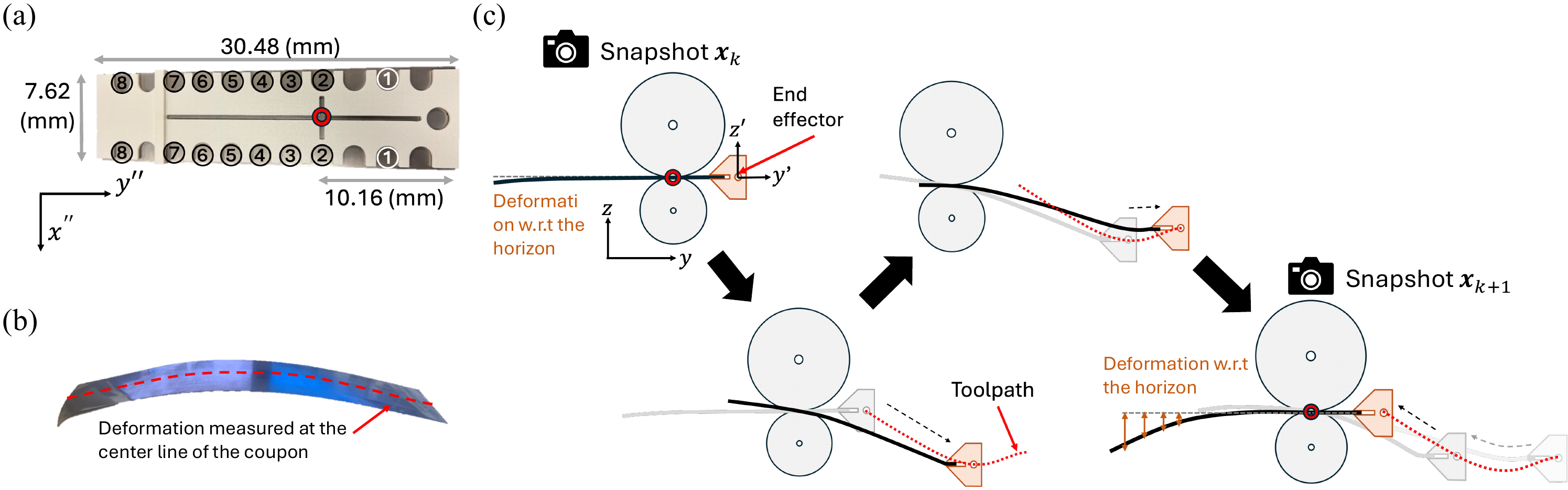}
    \caption{(a) Vicon tracker locations on the blank sheet, the red circle indicates the contact point with the wheel at the beginning and the end for each cycle. (b) An example of the final workpiece. (c) Illustration of a cycle of the forming process when one toolpath is applied from the side view. The red dotted curve is the toolpath represented by the trajectory for the end effector to track.}
    \label{fig:cycle}
\end{figure}

The integrated robotic English wheel system is illustrated in Figure \ref{fig:overall_flowchart}(b). The selected target geometry for this study is a curved thin strip, in which a prescribed midline profile is desired. A representative formed strip is shown in Figure \ref{fig:cycle}(b). All experiments were conducted using 316L stainless steel blanks measuring 30.48 mm × 7.62 mm × 0.06 mm in thickness.

During the manufacturing process, the UR robot actuates the sheet in the $y$–$z$ plane through a compliant gripper that applies force at two contact points near the sheet edges. The toolpath is thus defined by prescribing the $y$- and $z$-displacements of the robot end-effector, as illustrated in Figure~\ref{fig:cycle}(c). One \emph{cycle} is defined as a complete forward and return motion of the gripper along this toolpath, returning both the robot and sheet to their reference positions. After each cycle, a \emph{snapshot} of the system state is captured by the Vicon tracking system to record the resulting sheet deformation. Although the Vicon system is capable of tracking continuous deformation during forming, snapshots are recorded only before and after each cycle for two primary reasons. First, the deformation evolution during gripper motion is strongly nonlinear and exhibit very different dyanmics due to continuous varying tool–sheet contact, making intermediate data difficult to model accurately. Second, when both sheet ends are fixed, the springback effect is suppressed, reducing representativeness of the measured deformation. Recording deformation after the gripper returns and one end on free standing captures the full elastic recovery, ensuring the measured $\mathbf{x}_k$ accurately reflects the stable post-forming shape and improves the fidelity of data-driven model identification. In the data collection stage, each experiment consists of $N=6$ forming cycles. 

Figure \ref{fig:cycle}(a) shows the metrology setup, where Vicon optical trackers are affixed to the sheet at eight designated locations labeled 1–8. The trackers at location 2 correspond to the reference $y$-position of the sheet, representing the initial contact point between the upper and lower wheels (highlighted as a red circle in Figure \ref{fig:cycle}(a) and Figure \ref{fig:cycle}(c)). These trackers are positioned 10.16 mm from the sheet edge. Three additional Vicon markers are mounted on the gripper and UR robot to monitor tool motion. Based on these markers, three coordinate systems are defined: (1) a \textbf{world coordinate system}, where $x$ and $y$ lie in-plane along the wheel rolling and transverse directions, respectively, and $z$ denotes the out-of-plane axis; (2) a \textbf{coupon coordinate system}, defined using trackers at location 1 and those on the gripper and UR robot, to establish the reference plane for deformation measurement; and (3) a \textbf{robot arm coordinate system}, defined solely by the trackers on the gripper and UR robot. Within the coupon coordinate system, the $x''$, $y''$, and $z''$ positions of the sheet’s Vicon trackers are extracted after each forming cycle. As deformation along the $x''$-direction is negligible, the tracker data at both sheet edges are averaged to obtain the midline deformation profile in the $y$–$z$ plane.

Viewing the process as a dynamical system, the state and input is defined as follows. The system state vector, denoted as $\mathbf{x}_k \in \mathbb{R}^{n_x \times 1}$, represents the measured sheet deformation after the $k$-th forming cycle relative to the initial flat configuration, 
\begin{equation}
    \mathbf{x}_k = [\delta_1,\dots,\delta_{n_x}]^\top,
\end{equation} 
where $\delta_i$ denotes the out-of-plane displacement of the $i$-th Vicon tracker along the midline of the sheet. In this work, $n_x = 8$, corresponding to eight spatial measurement points distributed along the midline in the $y$–$z$ plane. The control input, denoted as $\mathbf{u}_k \in \mathbb{R}^{n_u \times 1}$, defines the commanded gripper trajectory for the $k$-th forming cycle, parameterized as a discrete sequence of $y$- and $z$-displacements:
\begin{equation}
    \mathbf{u}_k = [u_1,\dots,u_{n_u}]^\top,
\end{equation}
where $n_u = 20$ corresponds to the number of sampled points along the toolpath. 

An examination of process repeatability indicates that the maximum replication error at the free end of the coupon is within ±1.5 mm. The details of experiment sampling, modeling, and control design will be elaborated in in Section \ref{sec: modeling} and \ref{sec:MPC}.

\section{Technical Background}

This section introduces the theoretical foundations underpinning this work, including the Koopman operator theory, POD, and MPC. To ensure consistency with standard conventions in the literature, the notations used in this section follow commonly adopted formulations and are independent of those used in the rest of the paper.

\subsection{Koopman Operator}

\begin{figure}[b]
    \centering
    \includegraphics[width=0.6\linewidth]{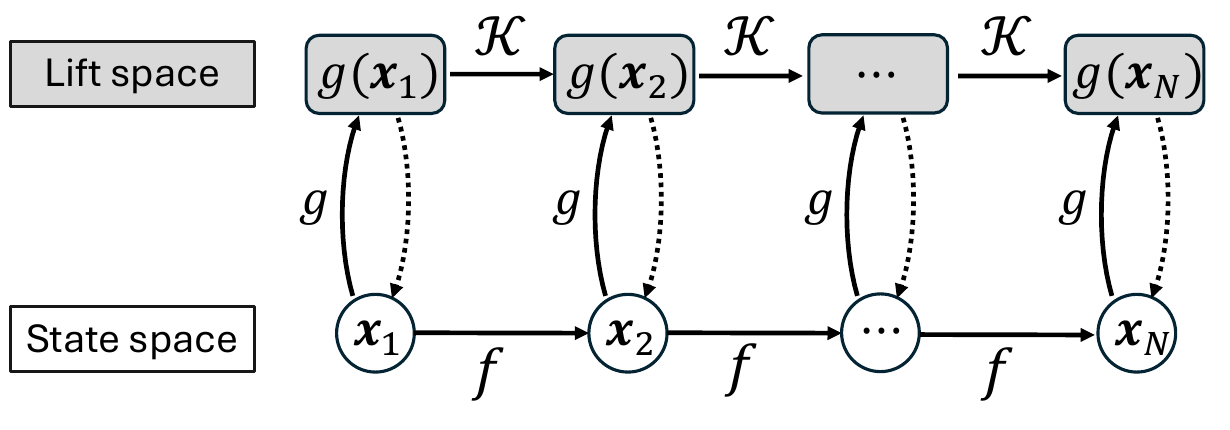}
    \caption{Illustration of Koopman operator theory. It shows that the evolution in the state space using nonlinear model $f$ is equivalent to the evolution of the observables in the lift space using linear observable $\mathcal{K}$. $g$ is a continuous, differentiable, and reversible function (diffeomorphism) that maps states to observables.}
    \label{fig:koopman theory}
\end{figure}

The Koopman operator theory proves the idea of representing a nonlinear dynamical system as a linear system by lifting the system states into an infinite-dimensional space of observable functions \cite{mauroy2020koopman}. Instead of directly modeling the nonlinear state evolution:
\begin{equation}
    \mathbf{x}_{k+1} = f(\mathbf{x}_k),
\end{equation}
the Koopman operator $\mathcal{K}$ acts linearly on the observables functions $g(\mathbf{x})$, such that
\begin{equation}
    \mathcal{K}g(\mathbf{x}_k) = g(f(\mathbf{x}_k)) = g(\mathbf{x}_{k+1}).
    \label{eq:koopman_evolve}
\end{equation}
This can be illustrated in Figure \ref{fig:koopman theory} that shows the relationship between $\mathcal{K}, f$, and $g$.  This formulation enables the use of linear analysis and control tools, such as eigenvalue decomposition, modal analysis, and linear MPC, to study and manipulate nonlinear systems, provided that an appropriate set of observables can be identified.

When control inputs are present, the system dynamics are extended to include the influence of control actions, written as
\begin{equation}
    \mathbf{x}_{k+1} = f(\mathbf{x}_k, \mathbf{u}_k),
    \label{eq:state-space}
\end{equation}
where $\mathbf{u}_k$ denotes the control input. The Koopman operator with control (often denoted as $\mathcal{K}_u$) generalizes the operator framework to capture the effect of $\mathbf{u}_k$ on the evolution of observables. A common formulation assumes separable dynamics between the state and control, leading to an affine linear representation in the lifted space \cite{brunton2019notes}:
\begin{equation}
    \mathbf{z}_{k+1} = \mathbf{A} \mathbf{z}_k + \mathbf{B} \mathbf{u}_k,
\end{equation}
where $\mathbf{z}_k = g(\mathbf{x}_k)$ is called the observables or the lifted state, and $\mathbf{A}, \mathbf{B}$ are finite-dimensional approximations of the Koopman operator and its control counterpart. This linear structure allows one to design and analyze linear controllers in the lifted space while maintaining interpretability with respect to the underlying nonlinear system.

The Koopman operator with control has attracted growing interest in data-driven modeling and control due to its ability to blend nonlinear system identification with linear control theory. Techniques such as eDMDc \cite{folkestad2020extended} and deep-learning-based Koopman architectures \cite{lusch2018deep} aim to identify $\mathbf{A}$ and $\mathbf{B}$ from data, offering a model that is both expressive and computationally efficient. When a control system can be represented as a linear state-space system, its corresponding MPC objective function can be formulated as a convex optimization problem. This problem can be solved using either the linear quadratic regulator (LQR) or quadratic programming (QP) method \cite{tsai2023robust}, depending on whether the problem is unconstrained or constrained. Both methods offer fast solution speeds and guarantee global optimality. This approach is particularly beneficial for real-time decision-making in DT, especially when dealing with nonlinear and complex problems where the MPC objective function is non-convex. In such cases, numerical solvers can be computationally expensive and prone to getting stuck in local optimums. 

\subsection{Proper Orthogonal Decomposition}

When identifying and modeling dynamical systems, the selection of system states plays a pivotal role in achieving effective dimensionality reduction and minimizing the need for extensive training data. In systems that exhibit pronounced spatial patterns, such as fluid flows \cite{ravindran2000reduced}, employing decomposition techniques enables the extraction of dominant spatial features. This approach allows the reduced-order model to capture and learn the temporal evolution of the system’s most critical spatial characteristics, thereby enhancing both modeling efficiency and physical interpretability.

Proper Orthogonal Decomposition (POD), also known as principal  Component Analysis (PCA) in statistics or the Karhunen–Loève expansion in stochastic processes \cite{chen2021unknown}, is a mature technique for extracting dominant modes from high-dimensional data. POD provides a low-dimensional representation of complex systems by identifying orthogonal basis functions that capture the most energetic or correlated structures in the data. 
It has become a cornerstone method in reduced-order modeling, flow analysis, and data-driven model identification for dynamical systems.

Consider a system state vector $\mathbf{x} \in \mathbb{R}^n$ with $m$ samples, forming a data matrix
\begin{equation}
    \mathbf{X} = [\mathbf{x}_1, \mathbf{x}_2, \dots, \mathbf{x}_m] \in \mathbb{R}^{n \times m},
\end{equation}
where each column represents a snapshot of the system state. The goal of PCA is to find an optimal set of orthonormal basis vectors $\{ \boldsymbol{\phi}_i \}_{i=1}^{r}$, known as the principal modes, that best approximate the snapshot data in the least-squares sense. This can be obtained by performing the Singular Value Decomposition (SVD) of the snapshot matrix:
\begin{equation}
    \mathbf{X} = \mathbf{U} \boldsymbol{\Sigma} \mathbf{V}^\top,
\end{equation}
where $\mathbf{U} = [\boldsymbol{\phi}_1, \boldsymbol{\phi}_2, \dots]$ are the principal modes, $\boldsymbol{\Sigma}$ contains the singular values, and $\mathbf{V}$ gives the PCA coefficients. The principal modes form an orthogonal basis ordered by their energy content (represented by its corresponding eigenvalue). By truncating the decomposition after $r$ dominant modes (where $r<<n$), one obtains a low-dimensional approximation:
\begin{equation}
    \mathbf{x} \approx \sum_{i=1}^{r} a_i\boldsymbol{\phi}_i(\mathbf{x}),
    \label{eq:PCA_decompose}
\end{equation}
which represents the states using a superposition of the essential dynamics $\boldsymbol{\phi}_i$ of the original system, weighted by PCA coefficients $a_i$, while reducing the input dimensions. Moreover, with the principal modes obtained, one can also infer the states by reconstructing the state vector given arbitrary PCA coefficients. 

% Formally, PCA seeks to minimize the reconstruction error
% \vspace{-15pt}\begin{equation}
%     \min_{\{\boldsymbol{\phi}_i\}} \sum_{k=1}^{m} \left\| \mathbf{x}_k - \sum_{i=1}^{r} a_i^{(k)} \boldsymbol{\phi}_i \right\|_2^2,
%     \vspace{-15pt}
% \end{equation}
% subject to orthonormality constraints $\boldsymbol{\phi}_i^\top \boldsymbol{\phi}_j = \delta_{ij}$, where $\delta_{ij}$ is a Kronecker delta. 

\subsection{Chebyshev Polynomials Decomposition}

A key limitation of PCA is that the reconstructed state vectors may lack smoothness, as only a subset of the principal modes is retained, and the modes themselves are not necessarily smooth. In applications where the reconstructed trajectories, such as the toolpaths of robotic arms, must exhibit smooth and continuous motion, additional post-processing of the PCA-reconstructed vectors is often required to ensure compliance with manufacturing or operational constraints.  

To address this limitation, one effective approach for functional decomposition of one-dimensional vector fields is the Chebyshev polynomial decomposition \cite{mason2002chebyshev}. Chebyshev polynomials form a family of predefined orthogonal basis functions derived from trigonometric identities of cosine and sine. The smoothness and completeness of the basis ensure that the reconstructed vector field is inherently smooth and differentiable, which makes it well suited to applications that require trajectory continuity or smooth gradient information.

%The Chebyshev polynomials of the first kind, denoted as $T_n(x)$, are defined over the bounded domain $x \in [-1, 1]$ by the recurrence relation:
% \vspace{-15pt}\begin{equation}
%     T_0(x) = 1, \quad T_1(x) = x, \quad 
%     T_{n+1}(x) = 2xT_n(x) - T_{n-1}(x), \quad n \ge 1.
% \end{equation}
% An equivalent trigonometric form is given by $T_n(x) = \cos(n \arccos x)$, which reveals their oscillatory nature and direct connection to Fourier-type representations.

% The Chebyshev polynomials are orthogonal with respect to the weight function $w(x) = (1 - x^2)^{-1/2}$:
% \vspace{-15pt}\begin{equation}
%     \int_{-1}^{1} T_m(x) T_n(x) w(x) \, dx =
%     \begin{cases}
%         0, & m \neq n, \\
%         \pi, & n = m = 0, \\
%         \pi/2, & n = m \neq 0.
%     \end{cases}
% \end{equation}

% This orthogonality minimizes redundancy among basis functions and leads to numerically stable decompositions with minimal interpolation error. Owing to these properties, Chebyshev polynomials are widely employed in approximating nonlinear functions and dynamical systems within bounded domains. 

In this work, the \textit{Chebyshev polynomials of the first kind} \cite{mason2002chebyshev} are employed, with the first few expressed as:
\begin{subequations}
\begin{align}
    T_0(x) & = 1, \\
    T_1(x) & = x, \\
    T_2(x) & = 2x^2 - 1, \\
    T_3(x) & = 4x^3 - 3x, \\
    T_4(x) & = 8x^4 - 8x^2 + 1, \\
    T_5(x) & = 16x^5 - 20x^3 + 5x.
\end{align}
\end{subequations}

Similar to PCA, a low-dimensional approximation can be achieved by truncating the Chebyshev series after $l$ terms, leading to:
\begin{equation}
    \mathbf{x} \approx \sum_{i=0}^{l} c_i T_i(\mathbf{x}),
    \label{eq:chebyshev_decomposition}
\end{equation}
where $c_i$ are the Chebyshev coefficients determined via orthogonal projection. Compared with PCA, however, the Chebyshev polynomial basis provides stronger guarantees of smoothness, better numerical conditioning, and superior approximation properties for one-dimensional smooth functions, thereby rendering it appropriate for smooth trajectory generation and reduced-order modeling of physical systems.

\begin{figure}[h!]
    \centering
    \includegraphics[width=0.75\linewidth]{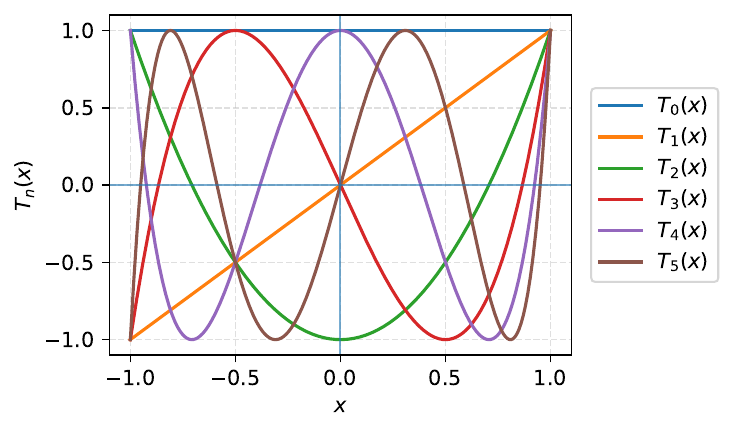}
    \caption{First six Chebyshev polynomials of the first kind, $T_0(x)$ – $T_5(x)$, over the interval $[-1, 1]$.}
    \label{fig:chebyshev}
\end{figure}

\subsection{Model Predictive Control (MPC) }

MPC is an advanced optimization-based control framework that utilizes a predictive model of the system to determine an optimal sequence of control actions over a finite time horizon~\cite{ tsai2023robust}. At each control step, MPC minimizes a performance cost that penalizes deviations from desired states and excessive control effort, subject to system dynamics and operational constraints. Only the first control input from the optimized sequence is implemented, and the optimization is resolved at the next time step with updated state measurements, which is a procedure commonly referred to as the receding horizon strategy.

Given the current state $\mathbf{x}_k$ and a prediction horizon $p$, the standard MPC problem can be formulated as follow to solve for optimal control sequence $\mathbf{u}$:
\begin{subequations}
\begin{align}
    \min_{\mathbf{u}=[\mathbf{u}_k,...,\mathbf{u}_{k+p-1}]} ~&J(\mathbf{u},\mathbf{x}_k)=\sum_{i=0}^{p-1}\left[||\mathbf{x}_{k+i}||_{\mathbf{Q}}^2+||\mathbf{u}_{k+i}||_{\mathbf{R}}^2\right], \label{eq:MPC_objective} \\
    s.t. \quad & \hat{\mathbf{x}}_{k+i+1}=\hat{f}(\mathbf{x}_{k+i},\mathbf{u}_{k+i}),~\forall i\in\mathbb{N}_{[0,p-1]}\label{eq:MPC_constraints_a}\\
       & \mathbf{x}_{k+i}\in\mathbb{X},~\forall i\in\mathbb{N}_{[0,p-1]},\label{eq:MPC_constraints_b}\\
      &  \mathbf{u}_{k+i}\in\mathbb{U},~\forall i\in\mathbb{N}_{[0,p-1]},\label{eq:MPC_constraints_c}
    \end{align}
    \label{eq:MPC_constraints}
\end{subequations} 

\noindent where $||\mathbf{x}||_{\mathbf{Q}}^2=\mathbf{x}^\top\mathbf{Q}\mathbf{x}$ represents quadratic penalty terms on the state and control vectors, respectively, with symmetric positive-definite weighting matrices $\mathbf{Q} \succ 0$ and $\mathbf{R} \succ 0$. The function $\hat{f}(\cdot)$ denotes the predictive model describing the system dynamics, while $\mathbb{X}$ and $\mathbb{U}$ define the admissible sets of states and control.

\section{Surrogate Modeling via Deep Koopman Operator}
\label{sec: modeling}
Due to the complex dynamics of the sheet metal forming process, it is impossible to iteratively simulate the process using physics-based finite element methods for real-time decision-making. Therefore, a surrogate model that can mimic the forming process is essential for online prediction and optimization through discrete event simulations (DES) for DTs. However, building a surrogate model for this process is extremely challenging as the evolution of the sheet deformation yields spatial and temporal behaviors, while collecting training data is also challenging due to experimental costs. In this section, we elaborate how we address these hurdles through the design of experiment (DOE), dimensional reduction, and eventually to train a deep Koopman operator. 

\subsection{Design of Experiments and Data Collection}

To generate smooth and executable toolpaths for the UR robot, the toolpath generator is formulated using a modified Fourier-inspired series with four design parameters that ensure trajectory continuity and manufacturability \cite{karkaria2024towards}. The toolpath, describing the $z$ location along the $y$ axis, is defined as:
\begin{equation}
\begin{split}
z(y) = A 
    \sin ( 2\pi fy + \phi) 
+\, Ty
\end{split}
\end{equation}
where $A\in[10,30]$, $f\in(0,1)$, and $\phi\in[0,2\pi]$ denote the base amplitude, frequency, and phase, and the parameters $T\in[-1,1]$ account for the linear trend. The range of these parameters are selected to ensure the generated toolpath follows a single mode curve without oscillation. 

This compact representation can flexibly synthesize one-dimensional periodic-based forming trajectories using low-dimensional parameterization, generating a variety of smooth and continuous toolpath that mimics the commonly seen, human-operated toolpath on the English wheel. A total of 20 base toolpaths were first generated within predefined parameter bounds—16 exhibiting negative curvature (downward, convex profiles) and four with positive curvature (upward, concave profiles). These base toolpaths were then systematically combined in sets of six using Latin Hypercube Sampling (LHS) to produce 80 composite toolpath sequences, each corresponding to a distinct forming plan. This LHS-based combinatorial strategy increases data diversity and ensures a broad coverage of the four-dimensional design space while maintaining trajectory smoothness and manufacturing feasibility.

To unify the representation of each fabrication process for a part, at the data collection stage, $N$ cycles are assigned to form a sequence of toolpaths. Therefore, the trajectory of the deformation field $\mathrm{\mathbf{X}}$ and the input toolpaths $\mathrm{\mathbf{U}}$ along the cycles for part $j$ is denoted as:
\begin{equation}
     \mathrm{\mathbf{X}}^j =  \begin{bmatrix}
\mid  & \mid  & \hdots & \mid \\
\mathbf{x}_0^j & \mathbf{x}_1^j & \hdots & \mathbf{x}_N^j \\
\mid  & \mid  & \hdots & \mid \\
\end{bmatrix}_{n_x\times (N+1)},  \mathrm{\mathbf{U}}^j =  \begin{bmatrix}
\mid  & \mid  & \hdots & \mid \\
\mathbf{u}_0^j & \mathbf{u}_2^j & \hdots & \mathbf{u}_{N-1}^j \\
\mid  & \mid  & \hdots & \mid \\
\end{bmatrix}_{n_u\times N}
\end{equation}
where $j \in [1,n]$, $n$ is the total number of experiments.

\subsection{Dimension Reduction}

The surrogate model in this framework is responsible for the prediction and design of the toolpath at each cycle to progressively realize a predefined target trajectory within a finite number of iterations via MPC. Hence, the model is constructed to predict the resulting sheet deformation after one complete cycle based on the current deformation state and the corresponding input toolpath, which is formulated as a state-space model as Equation \ref{eq:state-space}. Since $f$ is unknown, the purpose in this section is to introduce POD-based deep Koopman operator to identify a linear representation for the nonlinear $f$, allowing solving MPC in real-time using convex optimizations. 

To begin with, the deformation data for each part fabrication, denoted as $\mathrm{\mathbf{X}}^j$, are organized into two matrices that represent consecutive snapshots of the system:
\begin{equation}
    \mathrm{\mathbf{X}}_k^j =  \begin{bmatrix}
\mid  & \mid  & \hdots & \mid \\
\mathbf{x}_0^j & \mathbf{x}_1^j & \hdots & \mathbf{x}_{N-1}^j \\
\mid  & \mid  & \hdots & \mid \\
\end{bmatrix}_{n_x\times N}, \mathrm{\mathbf{X}}_{k+1}^j =  \begin{bmatrix}
\mid  & \mid  & \hdots & \mid \\
\mathbf{x}_1^j & \mathbf{x}_2^j & \hdots & \mathbf{x}_{N}^j \\
\mid  & \mid  & \hdots & \mid \\
\end{bmatrix}_{n_x\times N}. 
\end{equation}
% Thus, the concatenation of snapshots from all the experiments gives us:
% \vspace{-15pt}
% \begin{subequations}
% \begin{align}
%     \mathrm{\mathbf{X}}_k^{all} &= [\mathrm{\mathbf{X}}_k^1,\mathrm{\mathbf{X}}_k^2,\cdots,\mathrm{\mathbf{X}}_k^n]_{n_x\times (nN)}\\ \mathrm{\mathbf{X}}_k^{all} &= [\mathrm{\mathbf{X}}_{k+1}^1,\mathrm{\mathbf{X}}_{k+1}^2,\cdots,\mathrm{\mathbf{X}}_{k+1}^n]_{n_x\times (nN)}\\ \mathrm{\mathbf{U}}_k^{all} &= [\mathrm{\mathbf{U}}_k^1, \mathrm{\mathbf{U}}_k^2,\cdots,\mathrm{\mathbf{U}}_k^N]_{n_u\times (nN)}. 
% \end{align} \label{eq:raw_data_all}
% \end{subequations}
% \vspace{-15pt}

However, directly using $\mathrm{\mathbf{X}}_k$, $\mathrm{\mathbf{X}}_{k+1}$, and $\mathrm{\mathbf{U}}_k$ for system identification presents two major challenges. First, the identified model $\hat{f}$ would have high-dimensional inputs and outputs $\hat{f}:(n_u+n_x)\mapsto n_x$, which substantially increases the data requirements for accurate identification in a data-driven framework. Second, during the MPC stage, the optimized discretized toolpath generated from such a model may not be physically reasonable or directly executable. This necessitates the inclusion of additional constraints, such as smoothness penalties within each toolpath, to ensure feasibility. Nevertheless, maintaining convexity in the MPC formulation limits the expressiveness of these constraints, and linear penalties alone may be insufficient to enforce the desired smoothness of the optimized toolpaths.

To address these challenges, POD is employed to project the raw state and toolpath data into a reduced-order space. By extracting the principal modes that characterize the dominant spatial features of the system, POD enables a compact representation of both the deformation field and the toolpath. This transformation not only reduces the dimensionality of the data but also captures the intrinsic spatial correlations between deformation and toolpath, thereby facilitating their representation and design within the same reduced-order feature space. 

In this work, PCA is employed to decompose the deformation field into a linear combination of principal modes, with each mode weighted by its corresponding PCA coefficient. In contrast, the toolpath is represented using a linear combination of predefined Chebyshev polynomials through Chebyshev polynomial decomposition. With that, the dynamic model to be identified becomes a mapping between how the principal modes of the deformation evolve given some spatial feature of the toolpath. The motivation for using different decomposition methods lies in the distinct characteristics of the two domains. PCA’s variance-based decomposition is advantageous for system identification, as it isolates the most energetically significant modes of variation and preserves the dominant dynamics governing the system’s behavior. Conversely, since toolpaths inherently follow smooth polynomial trajectories, decomposition via Chebyshev polynomials ensures that the reconstructed paths remain smooth and continuous, thereby producing toolpath designs that satisfy manufacturability and operational constraints.

Following Equation \ref{eq:PCA_decompose}, applying PCA to each deformation field $\mathbf{x}$ gives:
\begin{equation}
    \mathbf{{x}}^j \approx \boldsymbol{\Phi}\mathbf{\tilde{x}}^j
\end{equation}
where $\mathbf{\tilde{x}}_j\in R^r$ is a vector of unique PCA coefficients, and $\mathbf{\Phi} = [\boldsymbol{\phi}_1(\mathbf{x}), \dots, \boldsymbol{\phi}_{r}(\mathbf{x})]$ is a truncated set of principal modes shared across all the deformation fields. Therefore, by representing $\mathbf{x}$ using $\mathbf{\tilde{x}}$, we can also denote $\mathbf{\tilde{X}}^j = [\mathbf{\tilde{x}}^j_0,\dots,\mathbf{\tilde{x}}^j_N]$.

%$\mathrm{\mathbf{\tilde{X}}}_k^{all} = [\mathrm{\mathbf{\tilde{X}}}_k^1,\mathrm{\mathbf{\tilde{X}}}_k^2,\cdots,\mathrm{\mathbf{\tilde{X}}}_k^n]_{r\times (nN)}$, and $\mathrm{\mathbf{\tilde{X}}}_{k+1}^{all} = [\mathrm{\mathbf{\tilde{X}}}_{k+1}^1,\mathrm{\mathbf{\tilde{X}}}_{k+1}^2,\cdots,\mathrm{\mathbf{{\tilde{X}}}}_{k+1}^n]_{r\times (nN)}$.

\begin{figure}[h!]
    \centering
    \includegraphics[width=0.5\linewidth]{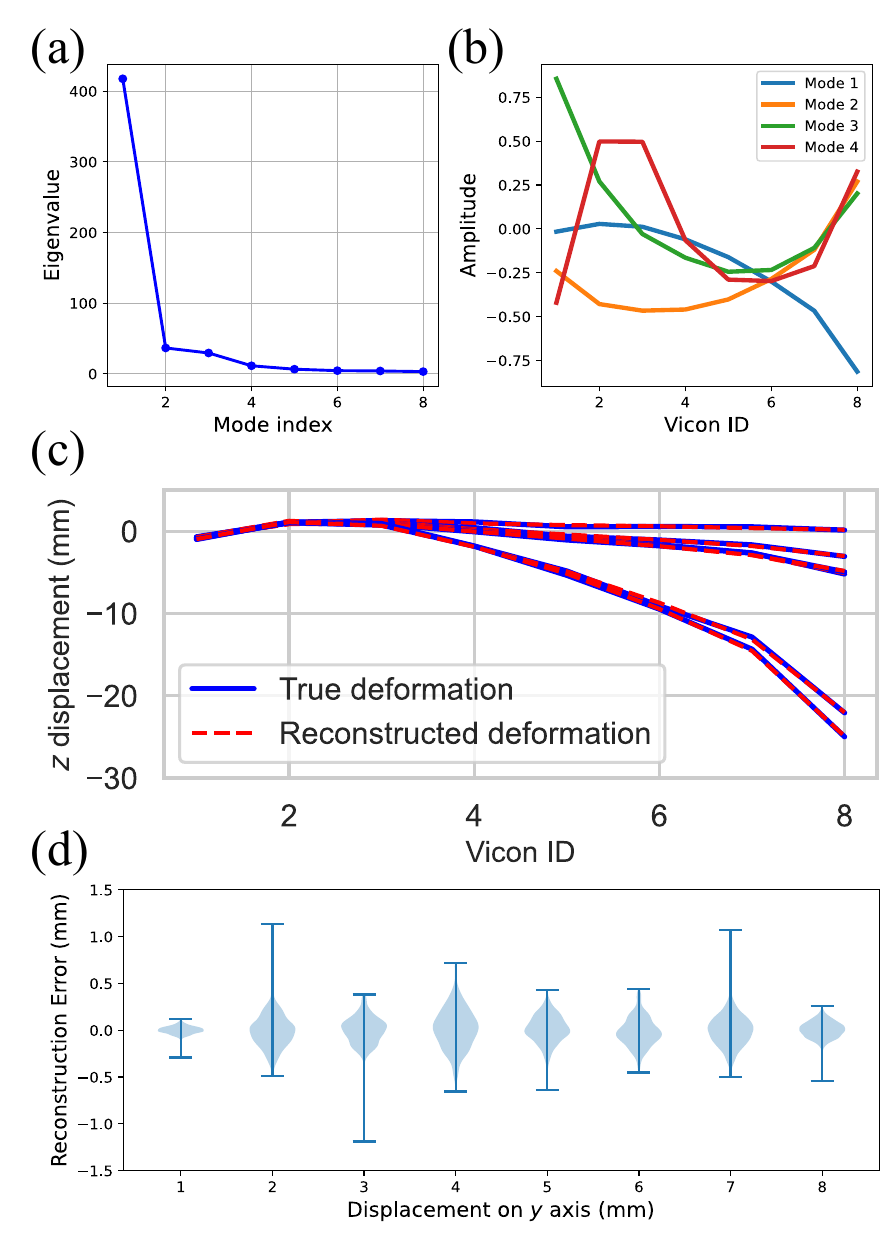}
    \caption{Dimensional reduction of sheet deformation using PCA. (a) The eigenvalue associated with each principal mode quantifies its contribution to the total variance, thereby indicating its importance. (b) Retained principal modes for this work. (c) Comparison between the reconstructed and the true deformation. (d) Reconstruction error on each Vicon location across all samples.}
    \label{fig:pca}
\end{figure}

The validation of the PCA-based dimensionality reduction for deformation analysis is presented in Figure \ref{fig:pca}. In this study, the first four principal modes (Figure \ref{fig:pca}(b)) are retained to reconstruct the sheet deformation. As shown in Figure \ref{fig:pca}(a), these four modes collectively capture over 99.5\% of the total variance, indicating that they effectively preserve the dominant deformation characteristics. The reconstructed deformation, illustrated in Figures \ref{fig:pca}(c) and \ref{fig:pca}(d), matches the true deformation, with reconstruction errors predominantly within $\pm 0.5$ mm.

Similarly, applying Chebyshev polynomial decomposition convert each toolpath $\mathbf{u}^j$ into a lower dimensional representation $\mathbf{\tilde{u}}^j$ following Equation \ref{eq:chebyshev_decomposition}:
\begin{equation}
    {\mathbf{u}}^j \approx \mathrm{\mathbf{T}}\mathrm{\mathbf{\tilde{u}}}^j
\end{equation}
where $\mathbf{\tilde{u}}^j\in R^p$ is a unique vector of the coefficients of the shared set of polynomials $\mathrm{\mathbf{T}}=[T_1(\mathbf{x}), T_2(\mathbf{x}),\dots,T_{r}(\mathbf{x})]$. Thus, the concatenated matrices using $\mathbf{\tilde{u}}$ are $\tilde{\mathrm{\mathbf{U}}}_k^j = [\mathbf{\tilde{u}}_0^j,\dots, \mathbf{\tilde{u}}^j_{N-1}]$.

%and $\mathrm{\mathbf{\tilde{U}}}_{k}^{all} = [\mathrm{\mathbf{\tilde{U}}}_{k}^1,\mathrm{\mathbf{\tilde{U}}}_{k}^2,\cdots,\mathrm{\mathbf{{\tilde{U}}}}_{k}^n]_{p\times (nN)}$. 

\begin{figure}[h!]
    \centering
    \includegraphics[width=0.5\linewidth]{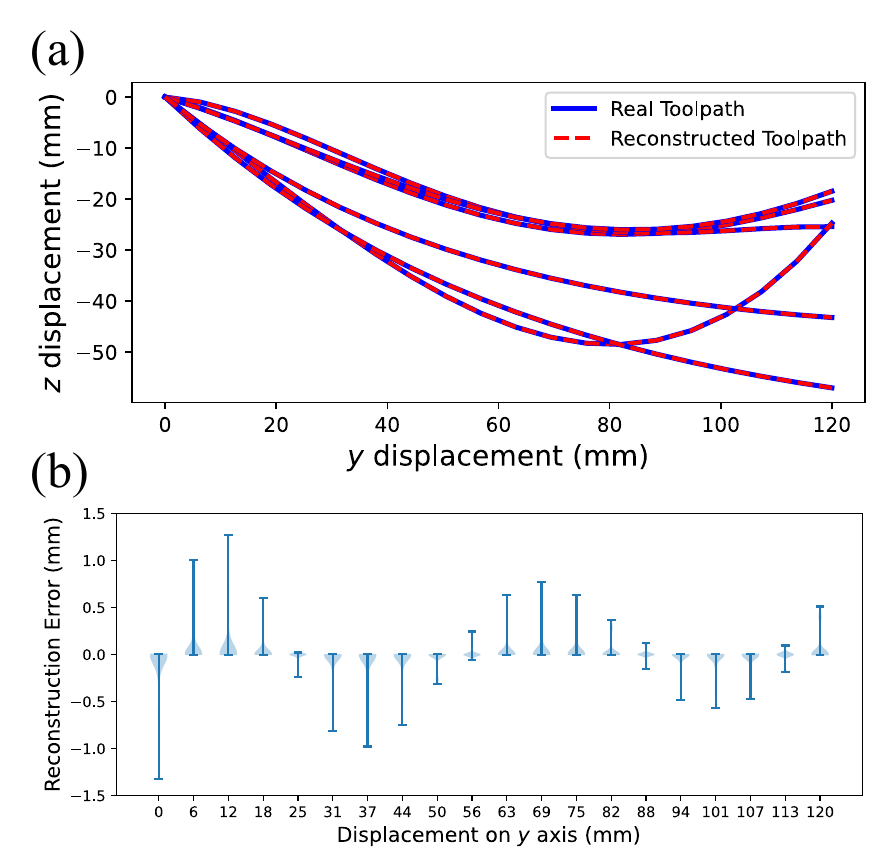}
    \caption{Dimensional reduction of toolpaths using Chebyshev polynomials. (a) Comparison between the reconstructed and the true toolpath. (d) Reconstruction error on each discrete point across all toolpaths.}
    \label{fig:cheby_reconst}
\end{figure}

Here, $p = 5$ basis functions ($T_0$ to $T_4$ in Figure \ref{fig:chebyshev}) are used to represent the toolpaths. As illustrated in Figure \ref{fig:cheby_reconst}(a), the reconstructed toolpath closely matches the measured (real) toolpath. The corresponding error distribution, shown in Figure \ref{fig:cheby_reconst}(b), indicates that most reconstruction errors lie within $\pm0.25$ mm, demonstrating the accuracy and effectiveness of the Chebyshev polynomial decomposition. Overall, we reduced the order of the system from $(8+20)\mapsto8$ to $(4+5)\mapsto4$ without compromising reconstruction accuracy. 

\subsection{POD-based Deep Koopman Operator}
With the reduced-order data, the objective of system identification is to learn the underlying dynamics represented as:
\begin{equation}
    \mathbf{\tilde{x}}_{k+1} = \tilde{f}(\mathbf{\tilde{x}}_k, \mathbf{\tilde{u}}_k).
\end{equation}

In this work, we adopt the deep Koopman operator framework \cite{lusch2018deep}, where a DNN learns the nonlinear lifting functions directly from data without prior specification. This approach enables the representation of highly nonlinear dynamics and allows dataset augmentation through temporal dependencies, thereby enhancing model generality and robustness. 

\begin{figure*}[t]
    \centering
    \includegraphics[width=1\linewidth]{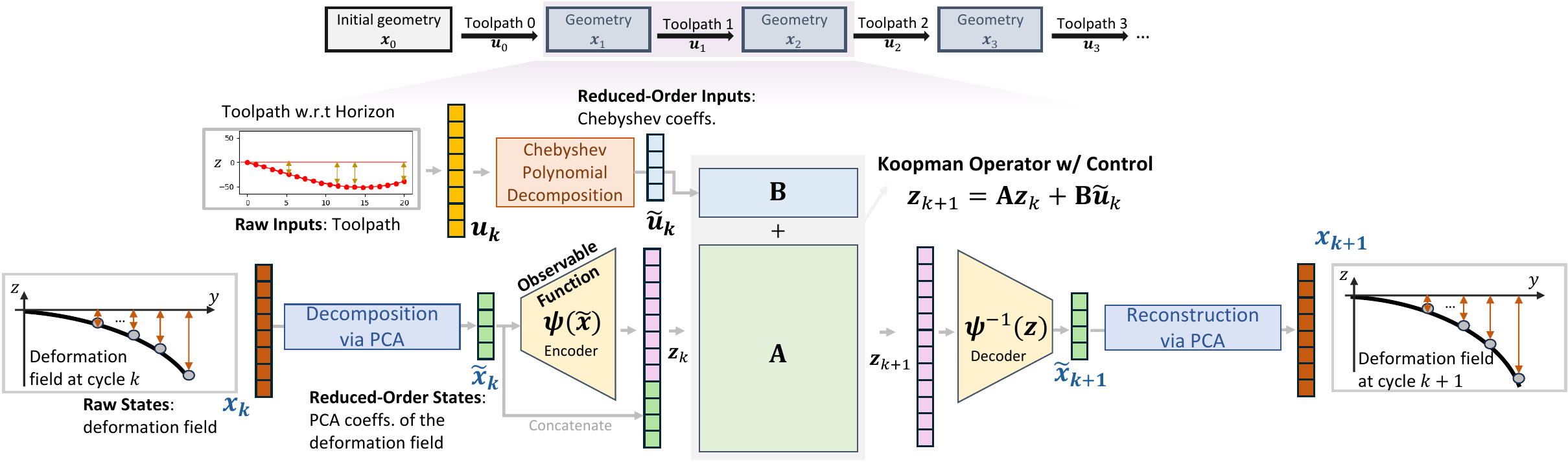}
    \caption{The proposed proper orthogonal decomposition (POD)-based deep Koopman operator for sheet metal forming.}
    \label{fig:Koopman}
\end{figure*}

The DNN architecture of the proposed POD-based deep Koopman operator with control for the sheet metal process is illustrated in Figure \ref{fig:Koopman}. It comprises three main components: an encoder, a linear evolution module, and a decoder. The fundamental goal of this setup is to address the question: \emph{How can we identify a finite-dimensional set of observable functions that best approximates the system’s nonlinear evolution in the linear lifted space?} The encoder maps the reduced-order states into a higher-dimensional lifted space $\mathbf{z}=\psi(\mathbf{\tilde{x}}) \in R^{d_z}$, serving as the learned observable functions. The linear matrices $\mathbf{A}$ and $\mathbf{B}$ then operate as the Koopman operator $\mathcal{K}$, governing the linear evolution of these observables. Finally, the decoder projects the lifted representation $\mathbf{z}$ back to the original reduced-order state space $\mathbf{\tilde{x}}$, ensuring a consistent reconstruction of the system dynamics. 

It is important to note that the reduced-order inputs are not mapped into the lifted space through nonlinear observables. This design choice is intentional: by maintaining the system states in a linear subspace, we preserve their interpretability and compatibility with subsequent MPC implementation. Lifting these inputs through nonlinear observables would compromise the linear structure required for efficient control synthesis, thereby limiting the applicability of linear MPC formulations.

\begin{table*}[h!]
\centering
\caption{Hyperparameters for network setup}
\small
\begin{tabular}{ccccccccc}
\hline \hline
hidden\_dim\_enc & hidden\_dim\_dec & batch size & epochs & lifted dim $d_z$ & dropout & scheduler  & $\lambda$   & steps ahead $S_p$ \\
{[}48,48{]}      & {[}48,48{]}      & 256        & 200    & 256 + 4        & 0.005   & *0.9 per 20 steps & 0.01 & 4           \\ \hline \hline
\end{tabular}
\end{table*}

Several training strategies are employed to improve data efficiency and model generality when identifying the Koopman operator under sparse data conditions. First, residual (skip) connections are incorporated into both the encoder and decoder to enhance training stability and increase the network’s nonlinear representational capacity, as suggested by \cite{szegedy2017inception}. Second, the reduced-order states are concatenated with the learned observables, enabling the Koopman framework to more effectively connect the linear dynamics in the lifted space with the underlying physical state \cite{brunton2016koopman}. Finally, dropout regularization is applied exclusively to the encoder and decoder to mitigate overfitting, while it is deliberately excluded from the linear matrices $\mathbf{A}$ and $\mathbf{B}$, which must remain deterministic during training to ensure consistent system identification. 

Leveraging $\mathbf{\tilde{X}}_k$ and $\mathbf{\tilde{U}}_k$ as training input and $\mathbf{\tilde{X}_{k+1}}$ as output, our loss function for training the deep Koopman operator has five weighted mean-squared error component: reconstruction loss $\mathcal{L}_{\text{recon}}$ on the encoder and decoder, one-step-ahead prediction loss $\mathcal{L}_{\text{one}}$, multi-step-ahead loss $\mathcal{L}_{\text{multi}}$ that penalizes prediction error with system rollout, stability loss $\mathcal{L}_{\text{stable}}$ penalizes cases where the eigenvalue of $\mathbf{A}$ is greater than 1 to force the identified system stable, and forward linearity loss $\mathcal{L}_{\text{lin}}$ that measures linearity consistency. Finally, an $l_2$ regularization on the weights $\mathbf{W}$ is added to avoid overfitting. More specifically:
\begin{subequations}
\begin{align}
    \mathcal{L} &= \alpha_1 \mathcal{L}_{\text{recon}} + \alpha_2 \mathcal{L}_{\text{one}} + \alpha_3 \mathcal{L}_{\text{multi}} + \alpha_4 \mathcal{L}_{\text{stable}} + \alpha_5 \mathcal{L}_{\text{lin}} + \lambda \left\|\mathbf{W} \right\|_2^2, \\
    \mathcal{L}_{\text{recon}} &= \|\mathbf{\tilde{x}}_k-\psi^{-1}(\psi(\mathbf{\tilde{x}}_k))\|_{\text{MSE}},\\
    \mathcal{L}_{\text{one}} &= \|\mathbf{\tilde{x}}_{k+1} - \phi^{-1}(\mathbf{A}\mathbf{z}_k+\mathbf{B}\mathbf{\tilde{u}}_k)\|_{\text{MSE}}, \\
    \mathcal{L}_{\text{multi}} &= \frac{1}{S_p}\sum_{m=0}^{S_p} \|\mathbf{\tilde{x}}_{m+1} - \psi^{-1}(\mathcal{R}\bigl(\psi(\mathbf{\tilde{x}}_{0}),\,\mathbf{\tilde{u}}_{0:m-1}\bigr))\|_{\text{MSE}}, \\
    \mathcal{L}_{\text{stable}} &=  \sum_{i=1}^{d_z} \text{ReLU}(\, \lambda_i(\mathbf A) - 1),  \\
    \mathcal{L}_{\mathrm{lin}} &= \frac{1}{S_p}\sum_{m=0}^{S_p} \|\psi(\mathbf{\tilde{x}}_{m+1}) - \mathcal{R}\bigl(\psi(\mathbf{\tilde{x}}_{0}),\,\mathbf{\tilde{u}}_{0:m-1}\bigr)\|_{\text{MSE}},
\end{align}
\end{subequations}

\noindent where $\mathcal{R}\bigl(\psi(\mathbf{\tilde{x}}_{0}),\,\mathbf{\tilde{u}}_{0:m-1}\bigr)$ is the recursive flow rollout given initial geometry $\mathbf{\tilde{x}}_0$ with multi-step control inputs, computed by $\mathbf{A}$ and $\mathbf{B}$, and $S_p$ is the rollout length for the multi-step-ahead loss. The weights are tuned via heuristic, where $\alpha_1=1000, \alpha_2=20, \alpha_3=10, \alpha_4 = 1000$, and $\alpha_5=5$. The deep Koopman operator is built with \texttt{Pytorch} and optimized with \texttt{Adam} with early stopping. The loss curve of the training process is in Figure \ref{fig:koopman_validation}(a). 

Overall, the proposed POD-based deep Koopman operator effectively bridges nonlinear feature learning and linear dynamical representation, enabling interpretable and computationally efficient system identification for the sheet-forming process.

\subsection{Model Validation}

\begin{figure}[h!]
    \centering
    \includegraphics[width=0.5\linewidth]{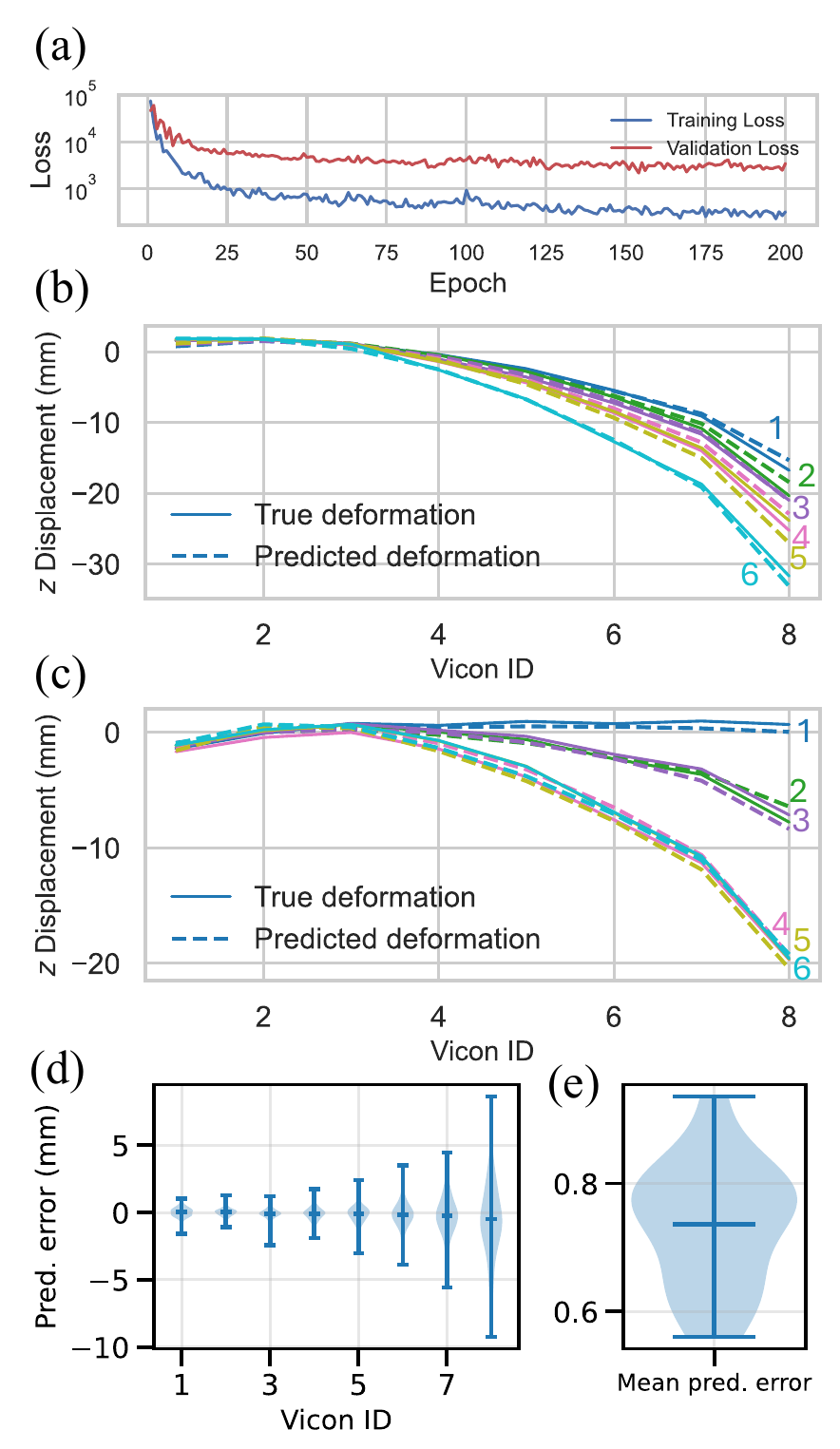}
    \caption{Validation of the learned Koopman operator. (a) Training and validation loss along epochs. (b) and (c) Comparison between predicted and ground truth sheet deformation after each cycle from two selected experiments. (d) prediction error distribution at each Vicon tracker location. (e) Distribution of average prediction error across all tracker location with 15 replicates.}
    \label{fig:koopman_validation}
\end{figure}

A total of 80 parts, each subjected to six forming cycles, were fabricated for model development. Among them, 75 experiments were used for training, while five were reserved for validation and testing. The model validation results are summarized in Figure \ref{fig:koopman_validation}. Figures \ref{fig:koopman_validation}(b) and \ref{fig:koopman_validation}(c) compare the true deformations from two representative validation experiments with the one-step-ahead predictions obtained from the learned Koopman operator. The predicted deformation trends and magnitudes show strong agreement with the measured data, confirming the model’s capability to capture the dominant dynamics of the forming process.

To evaluate model robustness, the training procedure was repeated 15 times with randomly shuffled datasets, each time leaving out a different subset for validation. Figure \ref{fig:koopman_validation}(d) presents the prediction error at each Vicon tracker location, aggregated across all replicates and validation experiments. The majority of prediction errors remain within $\pm5$ mm near the sheet tip, demonstrating the predictive accuracy of the learned Koopman operator. Figure \ref{fig:koopman_validation}(e) further summarizes the distribution of the mean absolute error (MAE) across all Vicon tracker locations using the validation sets, indicating that the learned operator maintains consistent performance regardless of data partitioning. These results collectively validate the accuracy and the stability of the proposed Koopman-based modeling framework. 

\section{Model Predictive Control via Koopman Operator}
\label{sec:MPC}
The MPC framework aims to design optimal toolpaths for the next $N = 6$ forming cycles to achieve the target geometry $\mathbf{r}$ with minimal iterations. Unlike conventional feedback controllers (e.g., PID), MPC explicitly incorporates multi-input multi-output (MIMO) system dynamics to prevent overshoot and iterative correction while enforcing constraints on both control inputs and predicted outputs for physically feasible solutions. After each forming cycle, the measured deformation is used to update the system state and reinitialize the optimization for the subsequent cycles, ensuring continual process optimality. To address material property variations over repeated cycles, an online model updating module is integrated into the MPC, allowing real-time refinement of the predictive model and improving control accuracy and adaptability.

\subsection{Model Predictive Control Formulation and Setup}

With the deep Koopman operator, the evolution of the deformation field under a given toolpath can now be represented using a linear state-space model. As the system dynamics is contracted in a linear representation, the MPC can be formulated as a convex optimization problem, allowing it to be solved using \texttt{qp}, which guarantees a global optimum and fast solving. 

The MPC for the sheet forming problem is formulated as:
\begin{subequations}
\begin{align}
    \min \quad &
    \sum_{k=0}^{N-1} \left[ (\mathbf{z}_k - \mathbf{z}_r)^\top \mathbf{Q} (\mathbf{z}_k - \mathbf{z}_r)
    + \mathbf{\tilde{u}}_k^\top \mathbf{R} \mathbf{\tilde{u}}_k \right] + (\mathbf{z}_P - \mathbf{z}_r)^\top \mathbf{Q}_N (\mathbf{z}_P - \mathbf{z}_r), \\
    \text{w.r.t.} \quad &{\mathbf{Z}=[\mathbf{z}_1,\dots,\mathbf{z}_{N}],\,\mathbf{\tilde{U}}}=[\mathbf{\tilde{u}}_0,\dots \mathbf{\tilde{u}}_{N-1}] \\
    \text{s.t.} \quad &
    h_1: \mathbf{z}_{k+1} = \mathbf{A}\mathbf{z}_k + \mathbf{B}\mathbf{\tilde{u}}_k, \quad k \in [0,N-1], \\
    & g_1: \mathbf{\tilde{u}}_{\min} \leq \mathbf{\tilde{u}}_k \leq \mathbf{\tilde{u}}_{\max}, \quad k \in [0,N-1],\\
    &  h_2: \mathbf{c}_1^\top{\mathbf{\tilde{u}}}_k = 0, \label{eq:constraint_1}, \quad k \in [0,N-1], \\
    & g_2: \mathbf{c}_2^\top {\mathbf{\tilde{u}}}_k < 0, \label{eq:constraint_2}, \quad k \in [0,N-1], \\
    & h_3: \mathbf{z}_0 = \psi(\mathbf{\tilde{x}}_k), 
\end{align} \label{eq:mpc_formulation}
\end{subequations}

\noindent where $\mathbf{z}_k$ denotes the lifted state in the Koopman space, $\mathbf{z}_r = \psi(\mathbf{r})$ is the target geometry mapped to the lifted space, and $\mathbf{\tilde{u}}_k$ represents the toolpath parameterized by Chebyshev coefficients. The weighting matrices $\mathbf{Q} = \mathrm{diag}([1,\dots,1,20,10,10,1]_{d_z})$ and $\mathbf{Q}_N = 0.1\mathbf{Q}$ penalize the accumulated and terminal deviations from the target geometry, respectively, while $\mathbf{R} = 10^{-5}\mathbf{I}_p$ regularizes the control effort to promote smooth actuation. The last four entries of $\mathrm{diag}(\mathbf{Q})$ correspond to the reduced-order states $\mathbf{\tilde{x}}_k$ that are concatenated with the lifted observables but not transformed by the nonlinear mapping. Assigning higher weights to these components prioritizes accurate tracking of the physically significant deformation features during optimization. The prediction horizon is set to $N = 6$, corresponding to six future forming cycles considered in each optimization. The MPC procedure terminates either after six forming cycles or when the maximum deviation between the target and current geometries falls below 1.5 mm.

The formulation above establishes a convex optimization problem that can be efficiently solved as a quadratic program (QP) using \texttt{osqp} \cite{osqp}. The linear equality constraint $h_1$ enforces the Koopman dynamics in the lifted space, ensuring consistency between the predicted and propagated states, while the inequality constraint $g_1$ limits the amplitude of the control input to prevent physically infeasible toolpaths. The boundary constraints $h_2$ and $g_2$ incorporate the properties of the Chebyshev representation into the optimization, guaranteeing geometric smoothness and ensuring that the designed toolpaths begin and end at physically consistent positions. Specifically, constraint $h_2$ enforces the toolpath to start at $z = 0$, where $\mathbf{c}_1$ is an alternating-sign vector defined as $\mathbf{c}_{1,i} = (-1)^{i+1}$, and constraint $g_2$ ensures that the toolpath displacement at $y = 120$ is smaller than that at $z = 0$, where $\mathbf{c}_2 = \mathbf{1}^\top$. Both constraints are derived from the properties of Chebyshev polynomials (see Figure~\ref{eq:chebyshev_decomposition}), where the polynomial values are $\pm1$ at $x = -1$ and equal to $1$ at $x = 1$.

The resulting sparse QP is implemented using the \texttt{OSQP} solver, which exploits the problem’s convexity and sparsity to achieve real-time computational performance. the average solving time for Equation \ref{eq:mpc_formulation}, which includes 1590 design variables, is 0.8067 seconds using an AMD 3975WX 32-Cores CPU.  In each cycle step, only the first toolpath is applied to the forming system following the receding-horizon principle. After each forming cycle, the measured deformation is used to update the current state $\mathbf{z}_0$, and the optimization is repeated to generate new toolpaths for the subsequent cycles.

\subsection{Online Model Updating}
A limitation of the Koopman representation trained on a small dataset is its restricted expressiveness in capturing global system dynamics. In particular, strain‐hardening effects are difficult to model because they depend on the material’s deformation history once the metal enters the plastic regime. As a result, the identified Koopman operator represents an average system response across varying material conditions without explicit dependence on material state. Given the limited training data, material state was not included as an additional input. Thus, an online model updating mechanism was implemented to adapt the toolpath in real time and mitigate model mismatch.

We employ a recursive least-squares (RLS) algorithm \cite{calderon2021koopman} to update only the control matrix $\mathbf{B}$ of the Koopman operator:
$\mathbf{z}_{k+1} = \mathbf{A}\mathbf{z}_k + \mathbf{B}\mathbf{\tilde{u}}_k$. Given the measured post-cycle geometry $\mathbf{x}_{k+1}^*$, the corresponding lifted state $\mathbf{z}_{k+1}^*$ is obtained through the composite mapping $\psi(\mathbf{\tilde{x}}_{k}^*)$. By fixing $\mathbf{A}$, the update of $\mathbf{B}$ becomes a multi-output linear regression problem formulated as
\begin{equation}
\mathbf{e}_k = \mathbf{z}_{k+1}^* - \mathbf{A}\mathbf{z}_k = \mathbf{B}\mathbf{\tilde{u}}_k + \boldsymbol{\varepsilon}_k,
\end{equation}
where $\mathbf{e}_k$ represents the residual and $\boldsymbol{\varepsilon}_k$ is the prediction error. Then, by assuming that the inaccurate $\mathbf{B}$ is the only source that contributes to $\boldsymbol{\varepsilon}_k$, the RLS minimizes the exponentially weighted least-squares cost, which leads to the conditional recursive update equations:
\begin{align}
\mathbf{K}_k &= \frac{\mathbf{P}_k \mathbf{\tilde{u}}_k}{\lambda + \mathbf{\tilde{u}}_k^\top \mathbf{P}_k \mathbf{\tilde{u}}_k}, \\
\mathbf{P}_{k+1} &= \frac{1}{\lambda}
\left( \mathbf{P}_k - \frac{\mathbf{P}_k \mathbf{\tilde{u}}_k \mathbf{\tilde{u}}_k^\top \mathbf{P}_k}
{\lambda + \mathbf{\tilde{u}}_k^\top \mathbf{P}_k \mathbf{\tilde{u}}_k} \right), \\
\mathbf{B} &\leftarrow \mathbf{B} + (\mathbf{e}_k - \mathbf{B} \mathbf{\tilde{u}}_k)\mathbf{K}_k^\top, 
\end{align}

\noindent where $\mathbf{P}_k = \mathrm{diag}(0.001\times\mathbf{1}_{d_z})$ denotes the covariance matrix, $\mathbf{K}_k$ is the adaptive Kalman gain and $\lambda=0.9$ is the adaptive ratio.

In our control law, $\mathbf{B}$ is updated only when one of two conditions is met: 1)  the maximum deviation between the measured and predicted geometries is greater than 3 \text{mm};  or 2) the sheet becomes stagnant, defined as a maximum deformation of less than 3\text{mm} between two consecutive cycles. This online RLS adaptation refines $\mathbf{B}$ in real time to compensate for material or actuation drift, while keeping the identified state dynamics $\mathbf{A}$ fixed. This prevents overfitting the Koopman operator to limited feedback data, which could otherwise compromise its global predictive accuracy. The resulting framework completes the DT integration for the forming process, enabling synchronized physical–digital states, informed decision-making, and continuous online model updating.

\subsection{Experimental Validation}

Two experimental cases were conducted to validate the proposed method using target geometries with large and moderate deformations. The experiments also assess the effectiveness of the online update of the input matrix $\mathbf{B}$.

\begin{figure*}[h!]
    \centering
    \includegraphics[width=1\linewidth]{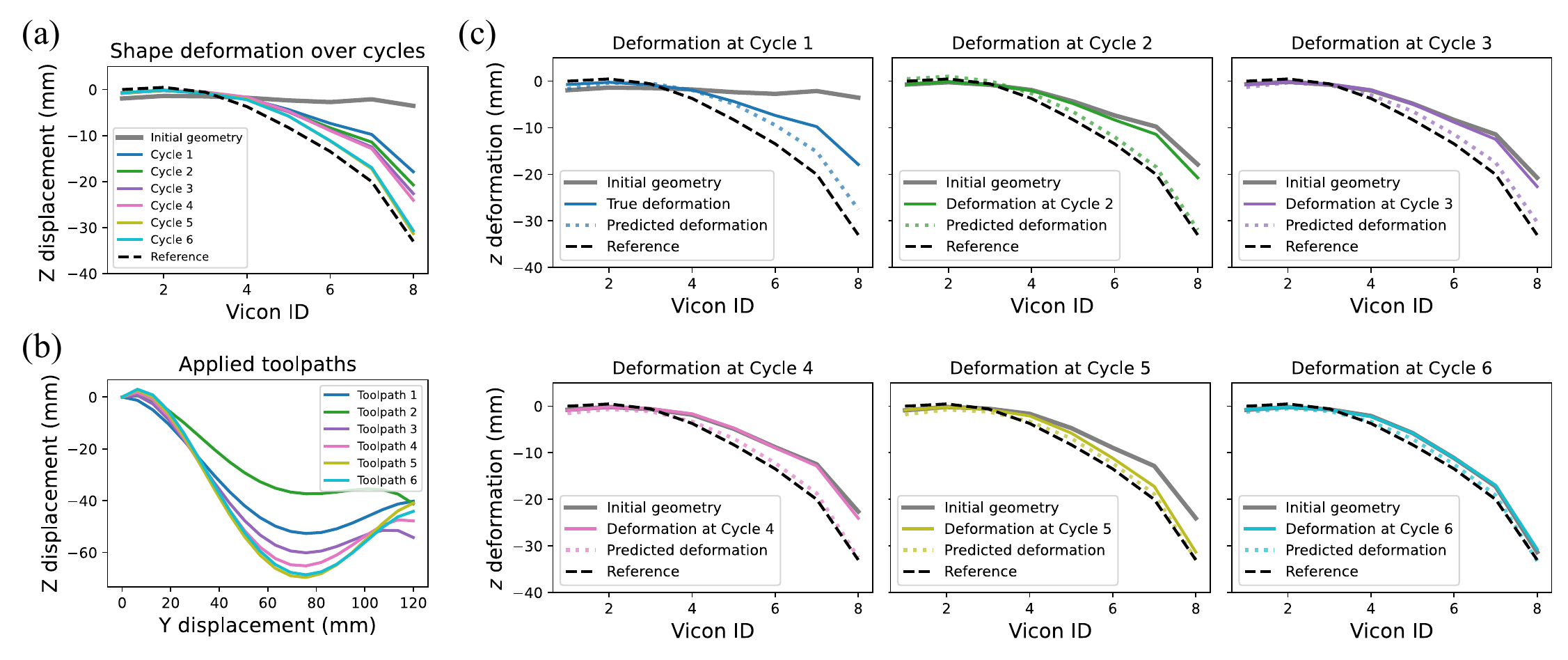}
    \caption{Result of sheet deformation over six cycles with model updating. (a) Sheet deformations over cycles. (b) Applied toolpaths over cycles. (c) Comparison of initial geometry, true deformation, and predicted deformation at each cycle.}
    \label{fig:exp1}
\end{figure*}

\begin{figure*}[h!]
    \centering
    \includegraphics[width=1\linewidth]{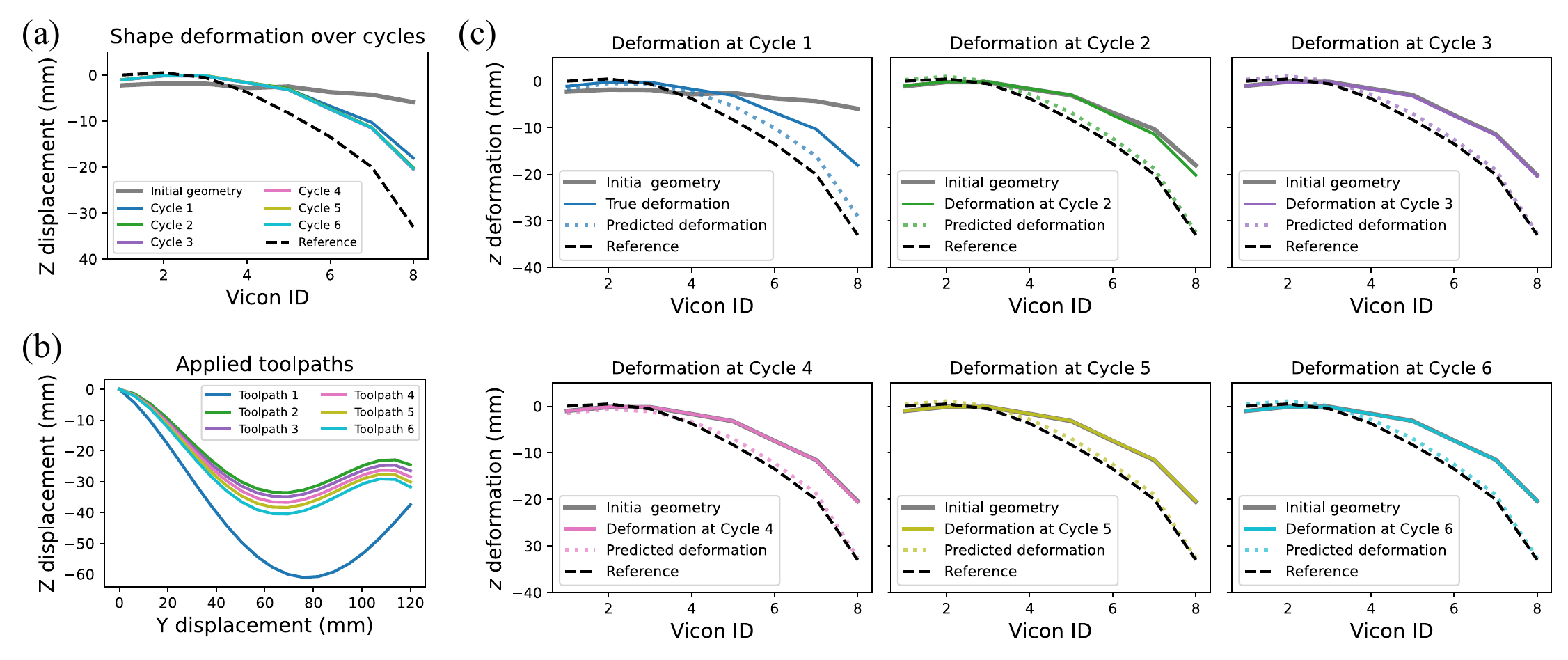}
    \caption{Result of sheet deformation over six cycles without model updating. (a) Sheet deformations over cycles. (b) Applied toolpaths over cycles. (c) Comparison of initial geometry, true deformation, and predicted deformation at each cycle.}
    \label{fig:exp1_wo}
\end{figure*}

\begin{figure}[h!]
    \centering
    \includegraphics[width=0.5\linewidth]{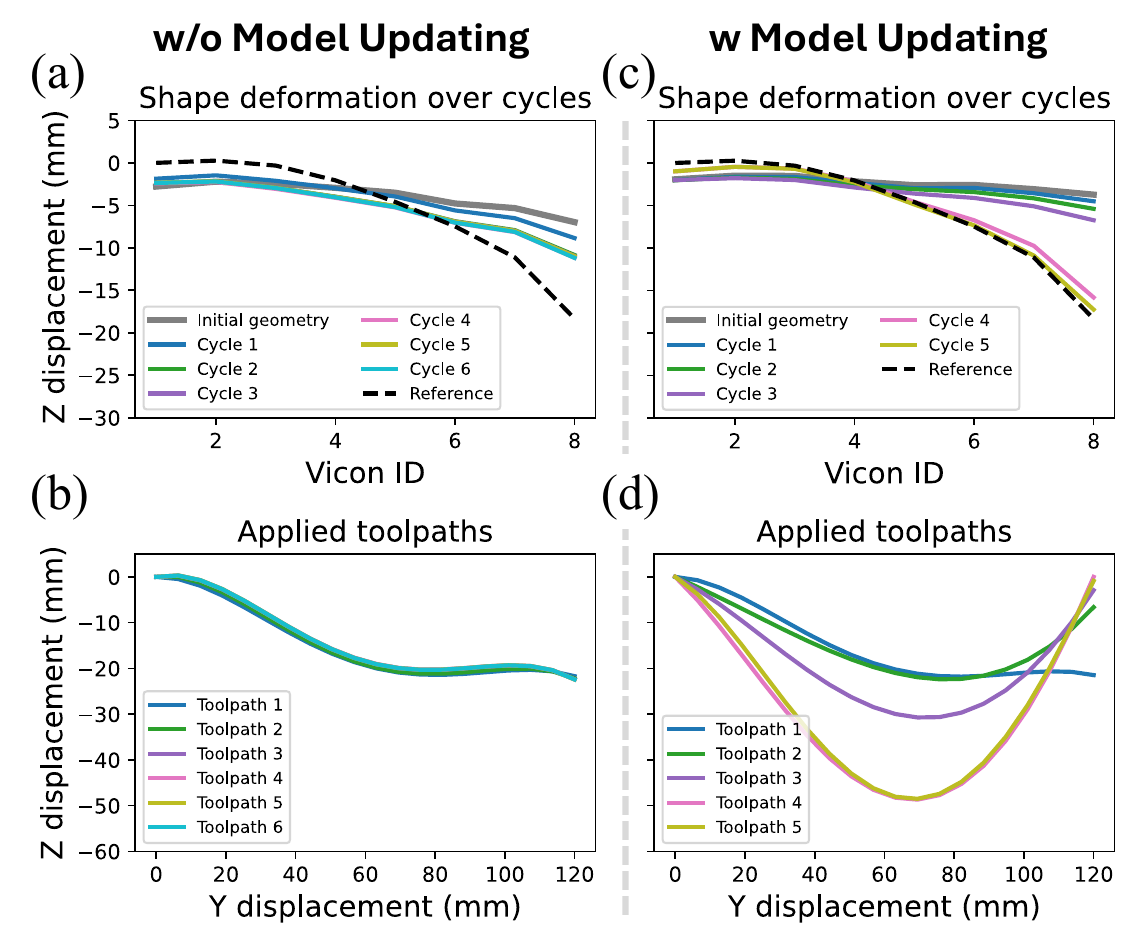}
    \caption{Result of sheet deformation given moderate-deformation target over five cycles without model updating. (a) Sheet deformation over cycles without model updating. (b) Applied toolpaths over cycles without model updating. (c) Sheet deformation over cycles with model updating. (d) Applied toolpaths over cycles with model updating.}
    \label{fig:exp2}
\end{figure}

Results for the large-deformation case are shown in Figures~\ref{fig:exp1} and~\ref{fig:exp1_wo}, corresponding to experiments with and without model updating, respectively. As shown in Figure~\ref{fig:exp1}(a), the sheet deformation progressively approaches the target geometry across cycles, while the applied toolpaths remain smooth owing to the Chebyshev-based parameterization. Without model updating (Figure~\ref{fig:exp1_wo}(c)), the model tends to overpredict the effect of each toolpath, leading to underbending and stagnation of deformation from Cycle~3 onward. As a result, the maximum difference between the final shape and the target is 12.64 mm at Vicon tracker location 8. This behavior arises because the nominal Koopman model overestimates toolpath effectiveness and cannot account for the evolving ductility of the material. When the online update of $\mathbf{B}$ is activated, the model adaptively corrects the toolpath gain based on prediction errors, generating steeper toolpaths in subsequent MPC optimizations and enabling sufficient bending to drive the sheet into the plastic deformation regime. This improves the maximum difference to 2.79 mm at tracker location 7. Similar trend can also be observed from the moderate-deformation case study, as shown in Figure \ref{fig:exp2}, where the maximum differences are 1.09 mm (reached terminating criteria after only five cycles) and 7.34 mm for cases with and without model updating, respectively. 

\begin{figure}[h!]
    \centering
    \includegraphics[width=0.7\linewidth]{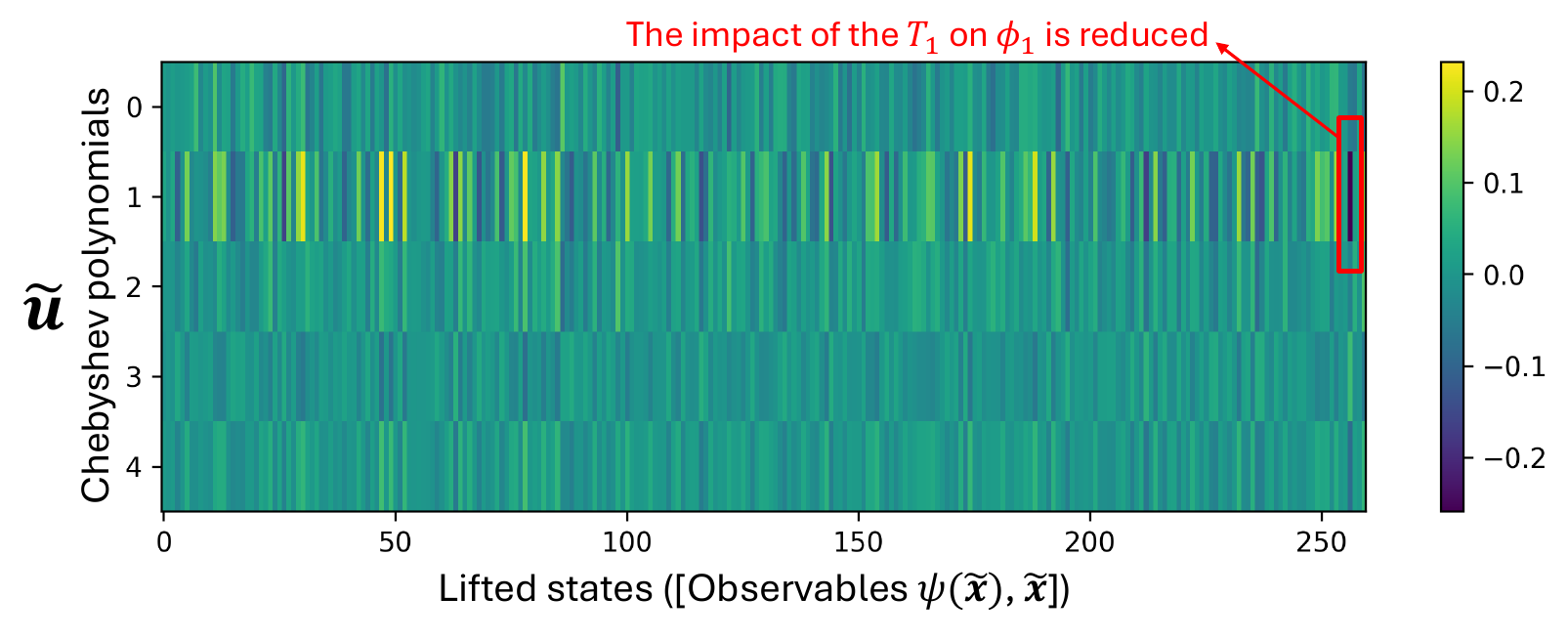}
    \caption{The increment of $\mathbf{B}$ matrix during model updating from one iteration.}
    \label{fig:B_inteprete}
\end{figure}

Figure~\ref{fig:B_inteprete} illustrates the change in $\mathbf{B}$ during one of the online updates, providing insight into the model adaptation mechanism. The most notable correction occurs in the gain associated with $\tilde{\mathbf{u}}(1)$, corresponding to the coefficient of the first-order Chebyshev polynomial $T_1$, which exhibits the largest magnitude adjustment among all control coefficients. Furthermore, the influence of $\tilde{\mathbf{u}}(1)$ on $\tilde{\mathbf{x}}(0)$, i.e. the coefficient of the dominant principal mode $\boldsymbol{\phi}_0$ concatenated as an augmented state in the lifted space, is significantly reduced. This modification alters the control gains, indicating that the actual control input is less effective than initially identified, thereby prompting the MPC to increase the control effort to achieve the target geometry using the updated model. These observations not only verify the effectiveness of the online model updating scheme, but also demonstrate the interpretability of the Koopman operator as a physic-based, explainable surrogate model.

The current model is limited in accurately representing all material states, as it serves as an aggregate approximation spanning material behaviors from the elastic to plastic regimes. The model may also be overfitted to the training and validation data, leading to reduced generalization and poorer performance on unseen conditions. One contributing factor is the potential sparsity of training data across the joint state-input space, which may be insufficient for the Koopman operator to fully capture material state-dependent variations, as noted in \cite{lusch2018deep}. This limitation can be addressed through active learning strategies aimed at maximizing information gain \cite{abraham2019active} via analyzing the Fisher information matrix on the Koopman operator, alongside continuous model updating using streaming data. Importantly, even when the Koopman operator with online updates does not achieve global accuracy, it remains highly effective in guiding control laws and directing the system toward desired outcomes \cite{mamakoukas2019local, uchida2024model}. This observation underscores the adaptability and practicality of the DT-based approach in decision-making.

\section{Closure}

In this work, we present an adaptive DT framework that integrates an adaptive POD-based deep Koopman operator for identifying the spatio-temporal dynamics of sheet deformation under toolpath actuation, followed by MPC for decision-making. Experimental validation on a robotic English Wheel system demonstrated that the proposed framework achieves modeling of the spatial-temporal deformation evolution, and real-time designing under dynamically evolving process and material conditions. Compared with having a static models, the adaptive DT provides enhanced flexibility, while remaining interpretability and generalization. These results demonstrate a practical application of real-time DT for manufacturing systems. This research advances the scope of DT in modeling, controlling, and self-updating for manufacturing systems characterized by coupled spatial and temporal dynamics, where both inputs and outputs are vector fields, bring a new avenue for next-generation manufacturing. The proposed pipeline provides a generalizable foundation for not only for digitalizing deformation-based manufacturing processes, but can be readily extended to other manufacturing processes, including incremental forming, forging, or additive manufacturing, that yields spatial-temporal dynamics and requires real-time optimization in actionable time. 

Future work will expand the framework to accommodate more complex geometries, including full 3D sheet forming, and on developing efficient multi-modal data acquisition strategies. In particular, we aim to integrate knowledge extracted from literature and human demonstrations with simulation and experimental data to enhance model refinement and enable more intelligent, autonomous DT systems. These efforts will advance DT development by enabling knowledge preservation, reuse, and intelligent decision-making in manufacturing.

\section*{Acknowledgement}

We appreciate the grant support from the NSF
Engineering Research Center for HAMMER under Award Number EEC-2133630, and NASA's Center for In-space Manufacturing (CISM-R2) under Award Number 80NSSC24M0176. Yi-Ping Chen also appreciates the Taiwan-Northwestern Doctoral Scholarship from the Minstry of Education in Taiwan.

\bibliographystyle{asmems4}
\bibliography{main} \clearpage

\end{document}